\documentclass[prd,nofootinbib]{revtex4}
\usepackage[colorlinks=true,linkcolor=red,citecolor=blue]{hyperref}
\usepackage{amsmath}
\usepackage{amsfonts}
\usepackage{graphicx}
\usepackage{subfigure}
\usepackage{dcolumn}
\usepackage{bm}
\usepackage{booktabs}
\usepackage[utf8]{inputenc}
\usepackage{multirow}
\usepackage{graphicx,graphics,dcolumn,booktabs,bm}
\usepackage{longtable,lscape}
\usepackage{txfonts}
\usepackage{overpic}
\usepackage{amssymb}
\usepackage{indentfirst}
\usepackage{epsfig}
\usepackage{feynmf}   
\usepackage{epstopdf}   
\usepackage{slashed}  
\usepackage{color}
\usepackage[section]{placeins}

\newcommand{\nn}{\nonumber}
\newcommand{\beq}{\begin{equation}}
\newcommand{\eeq}{\end{equation}}
\newcommand{\beqa}{\begin{eqnarray}}
\newcommand{\eeqa}{\end{eqnarray}}

\newcommand{\Bbar}{\,\overline{\!B}{}}
\newcommand{\Dbar}{\,\overline{\!D}{}}
\newcommand{\Kbar}{\,\overline{\!K}{}}
\def\B0bar{\Bbar{}^0}
\def\D0bar{\Dbar{}^0}
\def\K0bar{\Kbar{}^0}

\def\Heff{\mathcal{H}_{\rm eff}}
\def\O{\mathcal{O}}
\def\cbar{\overline{c}}

\def\ellbar{\overline{\ell}}

\graphicspath{{ps}}

\def\gf#1{\langle g_{#1}\rangle}

\begin{document}
\title{Unbinned Angular Analysis of $B\to D^*(D\pi)\ell \nu_\ell$ and $C_{V_R}$\footnote{Presented at the 30th International Symposium on Lepton Photon Interactions at High Energies, hosted by the University of Manchester, 10-14 January 2022.}}

\author{Zhuo-Ran Huang\footnote{Speaker.}}
 \email{zhuoran.huang@apctp.org}
 \affiliation{Asia Pacific Center for Theoretical Physics, Pohang, 37673, Korea}
\author{Emi Kou}
 \email{kou@lal.in2p3.fr}
\affiliation{Université Paris-Saclay, CNRS/IN2P3, IJCLab, 91405 Orsay, France}

\author{Cai-Dian L\"u}
\email{lucd@ihep.ac.cn}
\author{Ru-Ying Tang}
\email{tangruying@ihep.ac.cn}
\affiliation{Institute of High Energy Physics, Chinese Academy of Sciences, Beijing 100049, China\\}
\affiliation{School of Physics, University of Chinese Academy of Sciences, Beijing 100049, China}
\begin{abstract}
A sensitivity study of the unbinned angular analysis of the $B\to D^*(D\pi)\ell \nu_\ell$ decay is presented. In the analysis, it is shown that the Wilson coefficient of the right-handed vector current, $C_{V_R}$, can be measured to precision of 2-4$\%$ using either the full set of normalised angular observables or the forward-backward asymmetry $A_{FB}$ of the charged lepton in 10 $w$ bins. Such angular measurements are independent of the $V_{cb}$ puzzle.
\end{abstract}

\maketitle

\section{\label{sec:1}Introduction}
In recent years, the $B\to D^{(*)}\ell\nu$ decays have gained considerable attention due to several discrepancies between the experimental measurements and the Standard Model (SM) predictions, including the $R(D^{(*)})$ anomalies and the $V_{cb}$ puzzle. At present, the combined experimental results on $R(D^{(*)})={\mathcal B(B\to D^{(*)}\tau\nu)\over \mathcal B(B\to D^{(*)}\ell\nu)}$ by BaBar, Belle and LHCb deviate from the SM prediction by 3-4$\sigma$~\cite{Bernlochner:2017jka,Bigi:2017jbd,Jaiswal:2017rve,Bordone:2019vic,Iguro:2020cpg,Cheung:2020sbq,Huang:2018nnq}, which gives a hint of the violation of the lepton flavour universality. The other long-standing puzzle related to $b\to c\ell\nu$ transition is the $V_{cb}$ puzzle. The value of the Cabibbo–Kobayashi–Maskawa (CKM) matrix element $V_{cb}$ extracted from the inclusive decay $B\to X_c\ell\nu$ and the exclusive decays $B\to D^{(*)}\ell\nu$ deviate from each other. For the inclusive decay, all final hadrons containing a c quark are summed over, and the decay rate can be computed to high precision in an expansion in $\alpha_s$ and 1/${m_{c , b}}$ with the help of heavy quark expansion (HQE), optical theorem and operator product expansion (OPE). In contrast, for the exclusive decays, specified final hadrons are considered, therefore the decay rates receive large theoretical uncertainties from the hadronic form factors, though they are easier to measure experimentally. Currently the tension in $V_{cb}$ is at the level of $3\sigma$~\cite{Bernlochner:2017jka,Bigi:2017jbd,Jaiswal:2017rve,Bordone:2019vic,Iguro:2020cpg,Crivellin:2014zpa}, and such a deviation can be induced by either the theoretical and experimental uncertainties or new physics beyond the Standard Model.

There are already analyses for the $B\to D^{(*)}\ell\nu$ decay by Belle~\cite{Belle:2018ezy,Belle:2017rcc} and BaBar~\cite{BaBar:2019vpl}. These existing analyses are binned analyses, which means the data are binned in each of the kinematic angles and the momentum transfer and the $\chi^2$ projected to each of these parameters is utilized.  Unlike the binned analysis, in the unbinned analysis~\cite{Huang:2021fuc}, the measured values are obtained by maximizing the likelihood function summed over events. In such a way the full angular information can be used, and the analysis is expected to be more sensitive to new physics effects.

One of the simplest extensions of the SM is the right-handed vector current, which can be obtained from the left-handed vector current, i.e. the V-A current present in the SM, by flipping the chirality of the quark anti-quark fields. In the presented work, we propose to do an unbinned angular analysis of $B\to D^*(D\pi)\ell\nu$ to measure the Wilson coefficient of the right-handed vector current. Such measurements are expected to be done at future B physics experiments such as Belle II~\cite{Kou:2018nap} with large statistics.
\section{\label{sec:2}Theoretical Framework}
In this section, we introduce the theoretical framework of the $B \to D^*(D\pi)\ell\nu_\ell$ decay. In the weak effective theory, the Hamiltonian in charge of the $B \to D^*\ell\nu_\ell$ decay with the left-handed and right-handed vector currents (assuming no right-handed neutrino) is
\begin{equation}
   \Heff ={4G_F \over \sqrt2} V_{cb}\left[ C_{V_L}\O_{V_L} + C_{V_R}\O_{V_R} \right] + \text{h.c.} \,,
      \label{eq:Ham}
\end{equation}
where $G_F$ is the Fermi constant, and $V_{cb}$ is the CKM matrix element. The left-handed and the right-handed vector operators are defined as
\begin{equation}
 \O_{V_L} = (\cbar_L \gamma^\mu b_L)(\ellbar_L \gamma_\mu \nu_{L}) \,,  \,\,\,
     \O_{V_R} = (\cbar_R \gamma^\mu b_R)(\ellbar_L \gamma_\mu \nu_{L}) \,,
   \label{eq:operators}
\end{equation}
and $C_{V_L}$ and $C_{V_R}$ are the Wilson coefficients corresponding to $\O_{V_L}$ and $\O_{V_R}$. For the SM case, we have $C_{V_L}=1$ and $C_{V_R}=0$, but in some NP models, e.g. the left-right symmetric model~\cite{Kou:2013gna}, non-zero $C_{V_R}$ can be induced.

For the $B \to D^*(D\pi)\ell\nu_\ell$ decay, the differential decay rate in terms of the momentum transfer and 3 kinematic angles can be expressed by 12 independent angular observables through the following expression~\cite{Bernlochner:2014ova}:
\begin{eqnarray}
 \label{eq:rateJ}
&&\frac{\text{d} \Gamma( \bar B \to D^{*} (\to D \pi) \, \ell^- \, \bar \nu_\ell)}{\text{d} w \, \text{d}\cos\theta_V \, \text{d}\cos\theta_{\ell} \, \text{d}\chi} \nn\\
&=& \frac{6m_Bm_{D^*}^2}{8(4\pi)^4}\sqrt{w^2-1}(1-2 \, w\, r+r^2)\, G_F^2 \, \left|V_{cb}\right|^2 \,\mathcal{B}(D^{*} \to D \pi) \nn\\
&&\times\Big\{ J_{1s} \sin^2\theta_V+J_{1c}\cos^2\theta_V +(J_{2s} \sin^2\theta_V + J_{2c}\cos^2\theta_V )\cos 2\theta_\ell  +J_3 \sin^2\theta_V\sin^2\theta_\ell\cos 2\chi \nn \\
&&~~ + J_4\sin 2\theta_V\sin 2\theta_\ell \cos\chi
+J_5 \sin 2\theta_V\sin\theta_\ell\cos\chi + (J_{6s} \sin^2\theta_V+J_{6c}\cos^2\theta_V)\cos\theta_\ell \nn \\
&&~~ +J_7 \sin 2\theta_V\sin\theta_\ell \sin\chi+J_8\sin 2\theta_V \sin 2\theta_\ell\sin\chi + J_9 \sin^2\theta_V\sin^2\theta_\ell \sin2\chi\Big\}\,.
\end{eqnarray}
where $\theta_V$ is the angle between $D^*$ and $D$ in the $D^*$ rest frame, $\theta_l$ is the angle between the charged lepton and the virtual W in the W rest frame and $\chi$ is the tilting angle between the hadron plane and the lepton plane. These 3 angles along with the dimensionless momentum transfer $w$ can describe the four body $B\to D^*(D\pi)\ell\nu$ decay. Moreover, the angular observables $J_i$ are functions of helicity amplitudes and Wilson coefficients, which in our case are $C_{V_L}$ and $C_{V_R}$. The helicity amplitudes, $H_+$, $H_-$ and $H_0$ can be further expressed by the hadronic form factors.

To describe the hadronic form factors in the full $q^2$ range, we need a parametrization for the form factors. In our analysis, we study the two most commonly used parametrizations for $B\to D^*$ form factors, i.e. the CLN parametrization~\cite{Caprini:1997mu} and the BGL parametrization~\cite{Boyd:1997kz}. The CLN parametrization is based on the heavy quark expansion. The form factors are related each other under heavy quark symmetry therefore the number of free parameters can be reduced. For $B\to D^*$ form factors we only need 4 parameters, including the axial-vector form factor at zero hadronic recoil $h_{A_1}(1)$, the slope parameter $\rho_{D^*}^2$ to extrapolate $h_{A_1}$, and two form factor ratios at zero recoil, $R_1(1)$ and $R_2(1)$. The other parametrization, the BGL parametrization is derived following the analytic properties of the form factors, and it is maximally model-independent but has more free parameters. In the BGL parametrization, by separating the Blaschke factors absorbing the subthreshold $B_c$ poles and the outer functions calculated via OPE, the form factors are expanded in the conformal variable $z$ and truncated at a certain order given $z\ll1$. The expansion coefficients $(a^g_{0,1,\cdots}, a^f_{0,1,\cdots}, a^{\mathcal F_1}_{1,2,\cdots})$ are the free parameters to fit, and we keep $\vec{v}=(a^g_{0}, a^f_{0}, a^f_{1}, a^{\mathcal F_1}_{1}, a^{\mathcal F_1}_{2})$ in the sensitivity study following \cite{Belle:2018ezy}.
\section{Unbinned Angular Analysis of $B\to D^*(D\pi)\ell\nu$}\label{sec:unbinned}
For the unbinned angular analysis, angular observables can be obtained by maximizing the log-likelihood function, which is the sum of the logarithm of the normalized probability density function (PDF). The log-likelihood function is given as follows:
\begin{equation}
\mathcal{L}(\vec{\gf{}})=\sum_{i=1}^{N} \ln \hat{f}_{\vec{\gf{}}} (e_i)\,, \label{eq:55}
\end{equation}
where $\hat{f}_{\vec{\gf{}}}$ is the normalized PDF, $e_i$ are the experimental events which are functions of the kinematic angles, and $\vec{\gf{}}$ is the full set of normalized angular observables. $N$ is the total event number, which for Belle analysis is $\sim$95k~\cite{Belle:2018ezy}. $\hat{f}_{\vec{\gf{}}}(\cos\theta_V, \cos\theta_\ell, \chi)$ can be obtained by normalizing the differential decay rate and expressed in terms of the normalized angular observables $g_i$ as:
\begin{align}
\hat{f}_{\vec{\gf{}}}(\cos\theta_V, \cos\theta_\ell, \chi)=\frac{9}{8\pi}\times&\Big\{\frac{1}{6}(1-3 \gf{1c} +2\gf{2s}+\gf{2c}) \sin^2\theta_V+\gf{1c} \cos^2\theta_V \nn \\
&+(\gf{2s} \sin^2\theta_V+\gf{2c}\cos^2\theta_V )\cos 2\theta_\ell +\gf{3} \sin^2\theta_V\sin^2\theta_\ell\cos 2\chi \nn \\
&+\gf{4}\sin 2\theta_V\sin 2\theta_\ell \cos\chi
+\gf{5} \sin 2\theta_V\sin\theta_\ell\cos\chi+(\gf{6s} \sin^2\theta_V+\gf{6c}\cos^2\theta_V)\cos\theta_\ell \nn \\
&+\gf{7} \sin 2\theta_V\sin\theta_\ell \sin\chi+\gf{8}\sin 2\theta_V \sin 2\theta_\ell\sin\chi+\gf{9} \sin^2\theta_V\sin^2\theta_\ell\sin2\chi\Big\}\,,
\end{align}
where $g_i$ can be obtained from the un-normalized angular observables $J_i$ by dividing the following normalization factor
\begin{equation}
\gf{i} \equiv \frac{\langle J_i^{\prime}\rangle }{6\langle J_{1s}^{\prime}\rangle+3\langle J_{1c}^{\prime}\rangle-2\langle J_{2s}^{\prime}\rangle-\langle J_{2c}^{\prime}\rangle}\,.
\end{equation}
where $J^{\prime}_i \equiv J_i\sqrt{w^2-1}(1-2wr+r^2)$.
\section{\label{sec:3} Sensitivity study of $C_{V_R}$}
To perform a sensitivity study of the unbinned angular analysis, we need to generate pseudodata of $g_i$ to do the fit. We generate 10 bins of $g_i$ using the form factors measured by Belle~\cite{Belle:2018ezy}, in both CLN and BGL parametrizations, and use the toy Monte-Carlo method to generate the covariance matrices of $g_i$. In the toy Monte-carlo simulation, we set the total event number to be 95k according to the Belle analysis so that we can get the expected precision of $C_{V_R}$ for an unbinned analysis at Belle. For each $w$ bin, the event number can be obtained by using the form factors fitted by Belle~\cite{Belle:2018ezy}. For the CLN parametrization, the fitted results of the form factors lead to the following numbers of events in 10 $w$ bins:
\begin{eqnarray}
&N_{\rm event} =(5306, 8934, 10525, 11241, 11392, 11132,10555, 9726, 8693, 7497)\,.&  \label{eq:29}
\end{eqnarray}

Similarly, the numbers of events generated using the fitted results for the BGL parametrization are as follows:
\begin{eqnarray}
&N_{\rm event}=(5239, 8868, 10500, 11264, 11455, 11217, 10638, 9776, 8676, 7368)\,.&
\end{eqnarray}

Using the generated data for $g_i$ parameters, we can fit the form factors and $C_{V_R}$. It should be noted that $C_{V_R}$ cannot be determined in the fit using w-bins of the decay rate without knowing $V_{cb}$ precisely, because $C_{V_R}$ and $V_{cb}$ are strongly correlated as they both directly impact the decay rate. That’s why we need to use angular observables to fit $C_{V_R}$. In addition to the angular part of $\chi^2$, we also use the available lattice data of $h_{A_1}(1)$ in our fit\footnote{In fact the data of $h_{A_1}(1)$ is only useful for BGL parametrization because in CLN parametrization $h_{A_1}$ gets cancelled.}, as done in the Belle analysis. But with these data the fit still doesn’t converge because $C_{V_R}$ is also highly dependent of the vector form factor. Therefore, in order to do a sensitivity study, we use the central values of vector form factors obtained by Belle and assign them the errors expected from lattice, namely $4\%$ error for $R_1(1)$ and $7\%$ error for $h_V(1)$. Therefore, the $\chi^2$ used in our fit has the following form
\begin{equation}
\chi^2(\vec{v})=\chi^2_{\rm angle}(\vec{v})+\chi^2_{\rm lattice}(\vec{v})\,.
\label{eq:chi2sum}
\end{equation}
where $\chi^2_{\rm angle}(\vec{v})$ represents the angular terms, and $\chi^2_{\rm lattice}(\vec{v})$ is the constraint from lattice QCD.

The angular part $\chi^2_{\rm angle}(\vec{v})$ can be written as
\begin{eqnarray}
&&\chi^2_{\rm angle}(\vec{v}) =\sum_{w-{\rm bin=1}}^{10}\Big[\sum_{ij}N_{\rm event}\hat{V}_{ij}^{-1}(\gf{i}^{\rm exp}-\langle g_i^{\rm th}(\vec{v}) \rangle )(\gf{j}^{\rm exp}-\langle g_j^{\rm th}(\vec{v}) \rangle)\Big]_{w-{\rm bin}}\,, \nn
\end{eqnarray}
where $\langle g_i^{\rm th}(\vec{v})\rangle$ and $\langle g_i^{\rm ex}(\vec{v})\rangle$ are respectively the theoretical expressions and the experimental data (or pseudodata generated from toy-Monte Carlo simulation) for $g_i$ parameters, and $\hat{V}$ are the (generated) covariance matrices of $\langle g_i^{\rm ex}(\vec{v})\rangle$.

The lattice terms have the following form
\begin{equation}
\chi^2_{\rm lattice}(v_i)=\left(\frac{v_i^{\rm lattice}-v_i}{\sigma_{v_i}^{\rm lattice}}\right)^2 \,,\label{eq:38}
\end{equation}
where for the CLN fit we have $v_1=R_1(1)$, and for the BGL fit we have $v_1=h_{A_1}(1)$ and $v_2=h_V(1)$.

For the CLN fit, $h_{A_1}(1)$ gets cancelled, and by assigning $4\%$ error to $R_1(1)$ as expected from lattice QCD~\cite{Kaneko:2019vkx}, we obtain the following results
\begin{eqnarray}
&\vec{v}=(\rho^2_{D^*}, R_1(1), R_2(1), C_{V_R}) =
({1.106,1.229,0.852,0})\,,  &\nn\\
&\sigma_{\vec{v}}=({3.177,0.049,0.018,0.021}) \,,&\nn\\
&\rho_{\vec{v}}=\left(
\begin{array}{cccc}
 1. & -0.016 & -0.763 & 0.095 \\
 -0.016 & 1. & 0.006 & -0.973 \\
 -0.763 & 0.006 & 1. & -0.117 \\
 0.095 & -0.973 & -0.117 & 1. \\
\end{array}
\right)\,.
&
\end{eqnarray}
We find although the slope parameter $\rho_{D^*}^2$ has a large error, $C_{V_R}$ can be determined to high precision which is about $2\%$. From the correlation matrix we clearly see that $R_1(1)$ and $C_{V_R}$ are strongly correlated, which further explains why we need the lattice value of $R_1(1)$ to determine $C_{V_R}$. The BGL case is similar as shown in Eq.~\eqref{eq:bglfit}. Assigning the $7\%$ error estimated by lattice~\cite{Kaneko:2019vkx} to $h_V(1)$ , although one of the form factor parameters, $a^f_1$ gets a large error, $C_{V_R}$ can be determined to precision less than $4\%$. Also, the vector form factor at zero recoil in the BGL parametrization, $a^g_0$, is highly correlated to $C_{V_R}$.
\begin{eqnarray}
\label{eq:bglfit}
&\vec{v}=({a}^f_{0}, {a}^f_{1}, {a}^{\mathcal F_1}_1, {a}^{\mathcal F_1}_2, {a}^g_{0}, C_{V_R}) \nn \\
&=(0.0132,0.0169,0.0070,-0.0852,0.0241,0.0024)\,,  &  \nn\\
&\sigma_{\vec{v}}=({0.0002,0.0109,0.0026,0.0352,0.0017,0.0379}
)\,, &\nn \\
&\rho_{\vec{v}}=\left(
\begin{array}{cccccc}
 1. & 0.022 & 0.039 & -0.035 & 0.000 & 0.189 \\
 0.022 & 1. & 0.860 & -0.351 & 0.000 & 0.316 \\
 0.039 & 0.860 & 1. & -0.762 & 0.000 & 0.283 \\
 -0.035 & -0.351 & -0.762 & 1. & 0.000 & -0.119 \\
 0.000 & 0.000 & 0.000 & 0.000 & 1. & -0.923 \\
 0.189 & 0.316 & 0.283 & -0.119 & -0.923 & 1. \\
\end{array}
\right)\,.
&
\end{eqnarray}
It should be noted that the central value of the fitted $C_{V_R}$ is zero for both CLN and BGL, because our pseudo data is based on the Belle 18 measurement, where only the SM was considered. Here what makes sense are the uncertainties rather than the central values. Only if in future experimental analyses the right-handed contribution is considered, it will be possible to know the central value of $C_{V_R}$.
\begin{figure}[htbp]
\begin{center}
\includegraphics[scale=0.4]{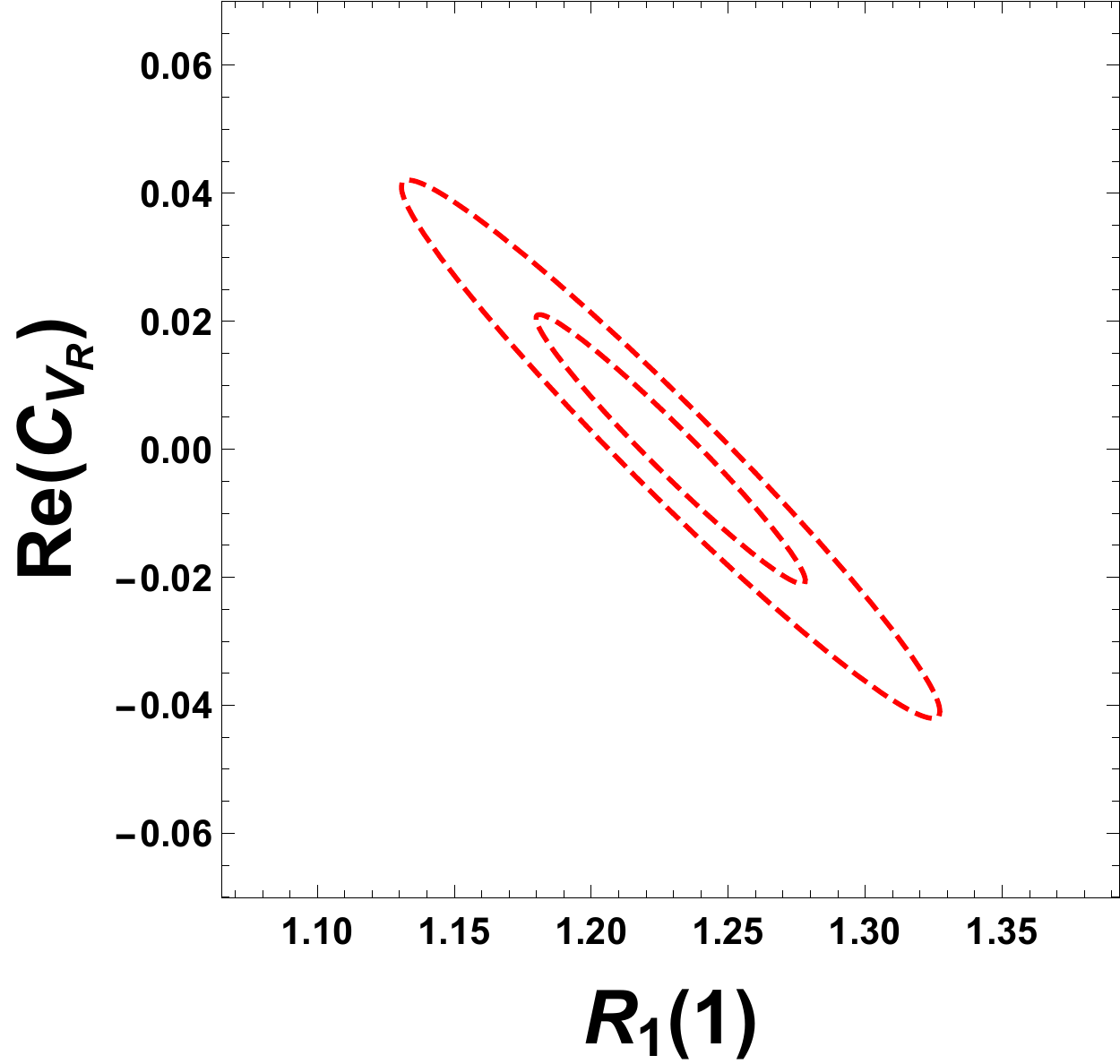}\hspace*{1cm}
\includegraphics[scale=0.4]{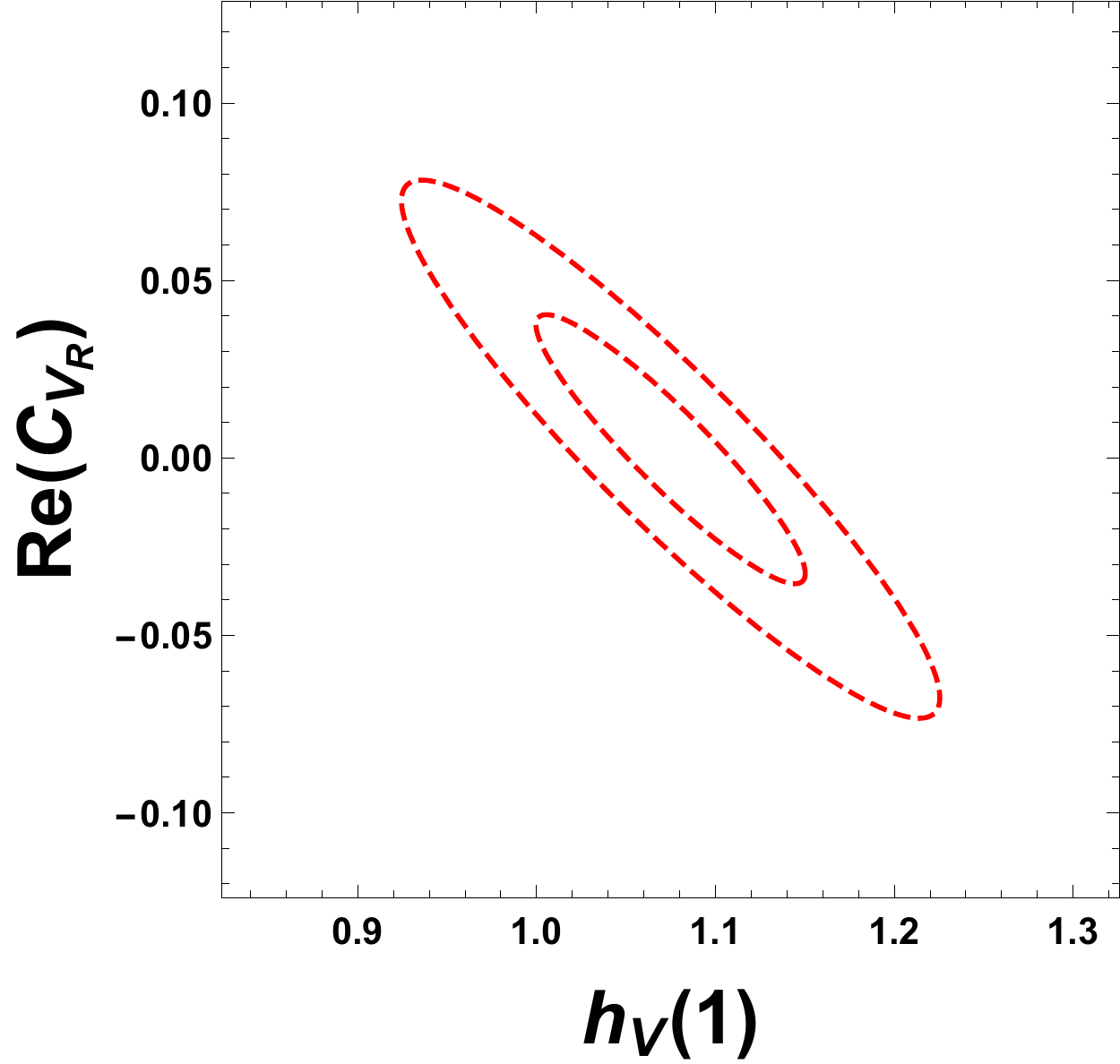}\hspace*{1cm}
\caption{$R_1(1)-C_{V_R}$ and $h_V(1)-C_{V_R}$ contours.}
\label{fig:countour}
\end{center}
\end{figure}
The correlation between $C_{V_R}$ and the vector form factor can be more clearly seen in Fig.~\ref{fig:countour}. The strong correlation between $C_{V_R}$ and the vector form factor shown by contour plots suggests if the lattice results\footnote{Lattice QCD calculation have been done recently for more $B\to D^*$ form factors respectively by Fermilab/MILC~\cite{FermilabLattice:2021cdg} and JLQCD~\cite{Kaneko:2021tlw}, although an official average is still needed.} turn out to be different from the measured value by Belle, the right-handed vector current can be a possible explanation.

In addition to the real part of $C_{V_R}$, we also study the imaginary part of $C_{V_R}$, and we find the imaginary part can be determined to precision of $0.7\%$ for both CLN and BGL parametrizations, which means the imaginary part of $C_{V_R}$ can be precisely determined from an unbinned angular analysis.

Besides the full angular analysis, we also study the role of the forward-backward asymmetry ($A_{FB}$) of the charged lepton in determining $C_{V_R}$. The forward-backward asymmetry turns out to be proportional to the angular observable $g_{6s}$ and it only requires one angle measurement. We still make $A_{FB}$ in 10 bins as done for the g parameters, and we find that for both the cases of CLN and BGL, $A_{FB}$ alone can determine $C_{V_R}$ to precision only slightly worse than that obtained by using the full set of $g_i$ parameters. Also, $C_{V_R}$ is highly correlated to the vector form factor, therefore we also need to know the vector form factor from lattice to determine $C_{V_R}$ by measuring $A_{FB}$. Since this single observable is particularly useful for constraining $C_{V_R}$, we highly recommend to measure it in the near future.
\section{\label{sec:6}Conclusions}
We performed a sensitivity study of the unbinned angular analysis of $B\to D^*(D\pi)\ell\nu$. We found the measurement of normalized angular observables $g_i$ are very useful for the determination of $C_{V_R}$ without the intervention of the $V_{cb}$ puzzle.

An important finding for the unbinned angular analysis is that $C_{V_R}$ and the vector form factor are highly correlated, therefore lattice input of the vector form factor would be crucial for the determination of $C_{V_R}$. In our sensitivity study, we found the real part of $C_{V_R}$ can be determined to precision of $2\%$-$4\%$ depending on the parametrization of form factors, and imaginary part of $C_{V_R}$ can be determined to precision of order $1\%$ in both CLN and BGL parametrizations. Having in mind that to explain $B\to D^*\ell\nu$ and the inclusive decay require $C_{V_R}$ to be $\sim5\%$, such precision of $C_{V_R}$ is already meaningful.

We also found that the forward-backward asymmetry $A_{FB}$ is very helpful for the determination of $C_{V_R}$. The single observable in ten $w$ bins can constrain $C_{V_R}$ to precision close to that achieved by measuring the full set of $g_i$ parameters, therefore we highly propose to do this single measurement in the near future.
\section*{Acknowledgments}
We would like to thank F. Le Diberder, T. Kaneko, P. Urquijo, D. Ferlewicz and E. Waheed for very helpful collaboration and discussions. This work was supported by TYL-FJPPL and by Natural Science Foundation of China under grants Nos 11521505, 12070131001 and National Key Research and Development Program of China under Contract No. 2020YFA0406400. Z.R. Huang acknowledges the YST program of the APCTP.
\bibliography{myreference}
\end{document}